# An Application of Transfer Hamiltonian Quantum Mechanics to Multi-Scale Modeling

Aditi Mallik[1], Carlos E. Taylor[2], Keith Runge[1] and James W. Dufty[1]

[1]*Department of Physics, University of Florida, FL-32611*
[2]*Department of Chemistry, University of Florida, FL-32611*

**Abstract** In quantum/classical (QM/CM) partitioning methods for multi-scale modeling, one is often forced to introduce uncontrolled phenomenological effects of the environment (CM) in the quantum (QM) domain as *ab initio* quantum calculations are computationally too intensive to be applied to the whole sample. We propose a method, in which two qualitatively different components of the information about the state of the CM region are incorporated into the QM calculations. First, pseudoatoms constructed to describe the chemistry of the nearest neighbor exchange interactions replace the atoms at the boundary of the CM and the QM regions. Second, the remaining effect of the CM bulk environment due to long-range Coulombic interactions is modeled in terms of dipoles. We have tested this partitioning method in a silica nanorod and a 3-membered silica ring for which *ab initio* quantum data for the whole system is available to assess the quality of the proposed partitioning method.

**Introduction**

Despite the latest computational advances, *ab initio* methods for studying complex phenomena like crack propagation, enzymatic reactions, stress corrosion, hydrolytic weakening of silica etc. are still far from routine. Such phenomena occur in macroscopic samples and quantum mechanical (QM) methods that include the effects of electron correlation are still limited to relatively small molecules. The alternative approach of classical molecular dynamics (MD) does not have this limitation, but an accurate description cannot be made using classical few body potentials, as they fail to work in states far from equilibrium or which involve breaking of chemical bonds or charge transfer. In such domains one requires QM methods to explicitly consider the electrons.

A promising tool for studying such phenomena is the application of *ab initio* QM techniques to small reactive regions (like the crack tip in case of studying crack propagation where bonds are breaking between atoms) and classical MD to describe the bulk (where the deformation is less). Such methods are referred to here as quantum mechanical / classical mechanical (QM/CM) methods. Over the past decade there has been intensive work developing such methods[1-9]. The first basic algorithm was laid by Warshel and Levitt[10].

One of the major problems with this approach is how to include the relevant effects of the bulk environment on the reactive region. There are two qualitatively different effects of the environment. The first is the short-range interaction at the QM/CM boundary where the partitioning involves cutting of covalent bonds between the two regions. Such a division results in a mismatch of geometry, forces, charge density etc., across the boundary of QM and CM regions. The second is the long range Coulomb interaction between the rest of the bulk and the QM region. As will be seen below, both effects must be accounted for to describe the QM region accurately. A successful QM/CM treatment will need a strategy to treat the bond cutting region, called termination schemes, and the longer range interactions through embedding.

There are various termination schemes for circumventing the problem of division of bonds. The most predominantly used scheme is the link atom (LA) method, first presented by Singh and Kollman[11]. In this method, hydrogen atoms are added to the CM side of broken covalent bond to satisfy the valency of the QM system. Interactions between link atoms and CM atoms are not included since link atoms do not have a physical existence in the system under study. Because of its simplicity, the LA method is still widely used in many types of QM/CM applications.

Another termination scheme is the frozen orbital method. Frozen orbital methods vary in the details of their implementations. In one of these methods, known as local self-consistent field (LSCF)[12], the electronic density along the frontier bond is represented by a strictly localized frozen atomic orbital, which has a preset geometry and electronic population. Reuter and co-workers[13] have compared the results for proton affinities and deprotonation enthalpies for a number of molecules using LA and LSCF methods and found the results to be comparable.

Gao *et al.* have proposed a refined LSCF method known as a generalized hybrid orbital (GHO) method[14] in which hybrid orbitals are divided into auxiliary and active orbitals, the latter of which are then optimized in the QM calculations. But both of these frozen orbital methods require substantial computational work.

There are numerous other termination schemes for the QM/CM boundary like the 'connecting bond' approach by Thiel *et al.*, 'pseudobond' scheme by Yang *et al.*, 'IMOMM' and 'ONIOM' procedures by Morokuma and co-workers[15], 'effective group potential (EGP)' by Poteau *et al*.[16] However these methods are not discussed here.

Here we report an approach, based on the application of the previously developed transfer Hamiltonian (TH)[17], which employs a newly parameterized pseudoatom to describe the short range interaction. The TH is a semi-empirical Hamiltonian which has been reparameterized to yield coupled cluster (CC) quality forces. MD is driven by the forces on the atoms, so that the TH is particularly trained to drive high quality MD results while requiring only the computational intensity of semi-empirical methods which are orders of magnitude less demanding than CC calculations on the same systems. We consider the application of pseudoatoms to a silica system. In the investigation of sample silica systems, the role of the remote environment becomes clear. The Coulomb interaction with the environment beyond the nearest neighbors is represented by its dipole to good accuracy, the subsystems having been chosen to be electrically neutral.

**Method**

We give only a brief introduction to the TH strategy as details are published elsewhere.[17] High level correlated quantum chemical calculations are very computationally intensive. The goal of the TH strategy is to provide forces for realistic MD simulations of the quality of the coupled cluster singles and doubles (CCSD) in computationally accessible times. To that end, a semi-empirical form of the Hamiltonian, (neglect of diatomic differential overlap-NDDO) has been reparameterized to reproduce CCSD forces from a representative small molecule. Pyrosilicic acid (Figure 1) has been used for the reparameterization of silicon and oxygen parameters and the results applied to silica systems.

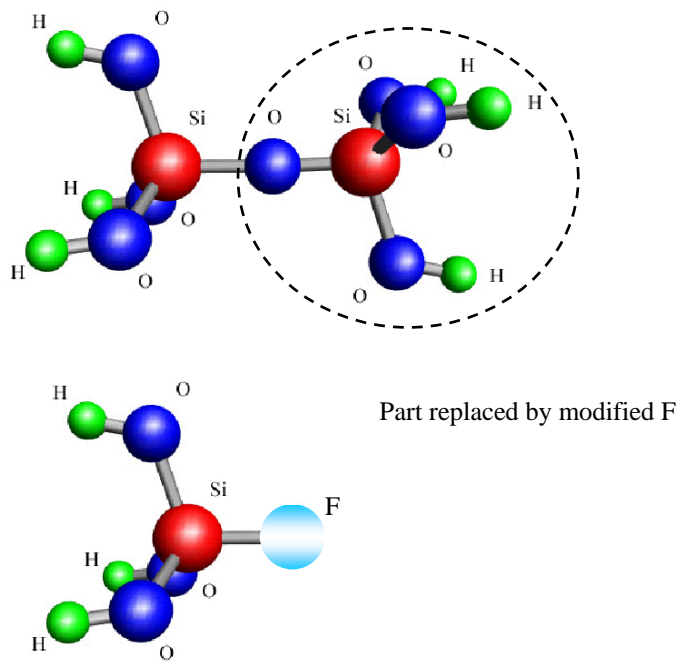

Figure 1: Training of Pseudoatoms on small system like Pyrosilisic acid

This TH strategy serves as the framework within which we investigate termination schemes for the embedding of a QM region inside a CM region. In the current application, we have presumed that the forces that arise from the TH are the reference forces by which the termination scheme is to be judged. A new embedding scheme is proposed for use in multi-scale simulations where the TH is used for the QM region, a pseudoatom is placed at the location of each boundary atom, and the electric dipole describes the remainder of the system.

In keeping with the TH strategy, we have based the parameterization of the pseudoatom by matching forces for the QM portion of the system. We intend that the short-range interactions, particularly the exchange interaction, will be taken into account using the pseudoatom, and we assume that they are transferable from the small model system to large systems. Beginning with NDDO parameters for fluorine, we modified these parameters to account for the near environment as based on a molecular model, pyrosilicic acid (Figure 1).

We include the long range interaction with the environment through its electric dipole. The long range interaction arises predominately from the Coulomb interaction of the environment with the QM region. The leading term in a multipole expansion of this long range effect is the electric dipole as the subsystem is taken to be electrically neutral. We confirm the sufficiency of the truncation of the multipole expansion at the dipole term in a system of modest size by comparing with the effect of the complete charge distribution.

We consider this partitioning problem in some benchmark systems for which a quantum treatment of the whole system is possible to assess the quality and physical basis for the proposed scheme. The results for two systems are described here: a silica nanorod, and a 3-membered silica ring associated with the surface structure of amorphous silica.

The silica nanorod, (developed by S. Yip et.al.)[18] contains 108 nuclei and is 16Å in length. It is made up of $SiO_2$ rings stacked one above another at an approximate distance of 3Å connected by oxygen atoms (Figure 2). It has been seen that "these nanorods are small enough to be authentic nanoscale objects and to be computationally tractable even for moderately refined QM approximations. Yet they are large enough to have authentic material properties e.g., stress strain curves and Young's modulus."[18] Moreover, the size of the system can be readily adjusted by the addition or the removal of rings.

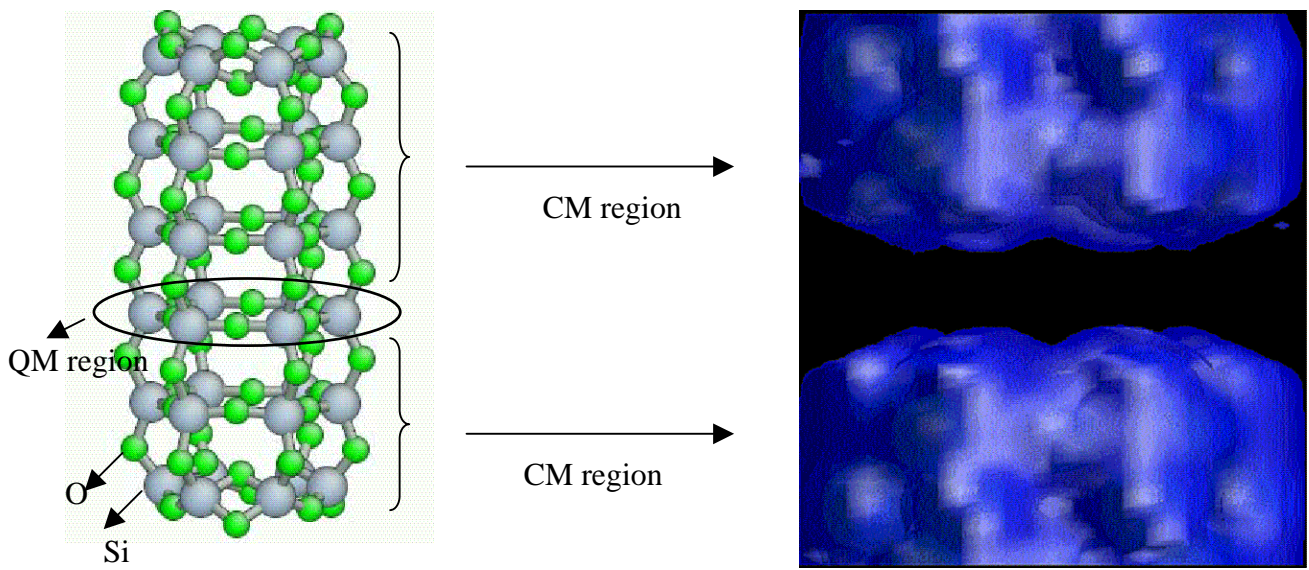

Figure 2(a): Partitioning of silica nanorod

Figure 2(b) Valence electron cloud of CM region

In this rod, one of the rings near the center of the rod is chosen to be the QM domain A and rest of the rod is treated classically [Figure 2(a)]. The localized nature of the valence electron charge density of the CM region B [Figure 2(b)] ensures the appropriateness of such a partitioning. The lighter shades represent electron higher densities, the black region in-between shows there is no overlap between the two clouds. It appears, therefore, that the partitioning we have chosen will allow a good first test of our newly developed termination scheme within the TH technique. The oxygen atoms that are bound to region A will be replaced with pseudoatoms and the remainder of region B will be represented by it electric dipole to account for it Coulomb interaction with region A.

We also present an analysis of the systems based on the reproduction of the charge density in the QM region. This test for the fidelity of a termination scheme is the subject of our current examination and a development of the theoretical basis of our approach will appear elsewhere.[19]

**Results**

Our analysis has three components. First, using the TH[17] - NDDO method[20], the forces and the charge density for the entire rod are calculated as the reference 'exact' QM description. In particular, it provides a description of the selected sub domain A with its correct chemical environment for the entire rod. Next, these forces and charge densities are calculated for the isolated QM region A with a simplified representation of the environment B: nearest neighbor pseudoatoms plus dipoles for the remainder of B above and below. For comparison we also study the choice of link atoms instead of pseudoatoms.

Our characterization of the partitioning method is based on the quality of the predicted charge densities and forces in the QM domain A. First, the charge density of the (i) ring with LA, and (ii) ring with pseudoatoms is compared to that of ring in bulk (i.e., the charge density for A resulting from the benchmark QM calculation for the entire rod). The density is calculated on a planar grid placed parallel to the plane of ring (which is taken to be at z=0). Visualization of the charge densities provides some insights into the descriptive differences between the pseudoatom and LA. Figure 3(a), Figure 3(b), Figure 3(c) and Figure 3(d) show the charge density of the isolated ring, the ring with LA, the ring with pseudoatoms, and the ring in bulk respectively in the plane of the ring. The six red spots (high density) on the contour of Figure 3(a) correspond to the six oxygen nuclei of the ring. The six blue spots (low density) in contour of Figure 3(b) correspond to the link atoms. In the contour of Figure 3(c), an overlap between the spots indicates the bonding between the nuclei similar to the ring in the bulk [Figure 3(d)]. Note that in Figure 3(d) the contour is projected on the top for clarity. These plots also indicate that pseudoatoms are a more realistic representation of the bulk compared to the link atoms because it takes into account the bonding of the ring. Table 1 shows the normalized difference of the square of the charge density in these two cases with respect to ring in bulk. Note that as the z coordinate gets larger we approach the pseudoatom location where we would not expect the termination scheme to reproduce the 'exact' result, hence we stop at z = 0.8 Angstroms. The right most column of the table shows that, while the difference is least for the ring with pseudoatoms, still the discrepancy from the 'exact' result is greater than 1% even in the best case and the relative agreement gets worse the further away from the ring we look. Hence, even though the pseudoatom provides reasonable chemical bonding behavior, there is still some impact from the rest of the system.

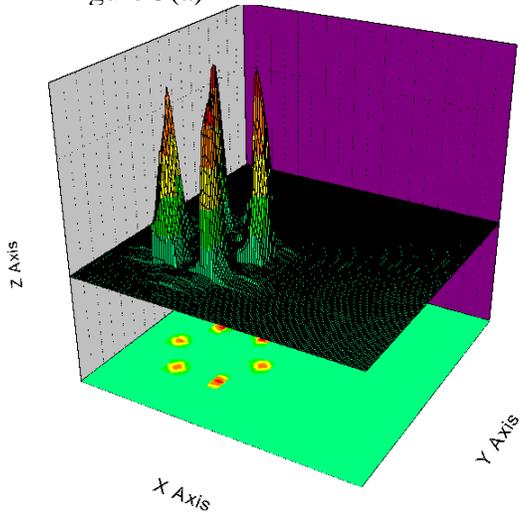

Figure 3(a)

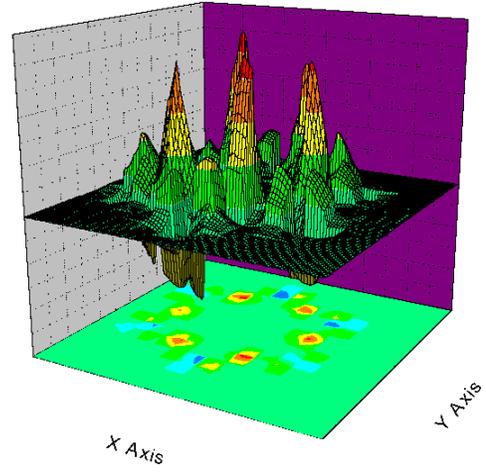

Figure 3(b)

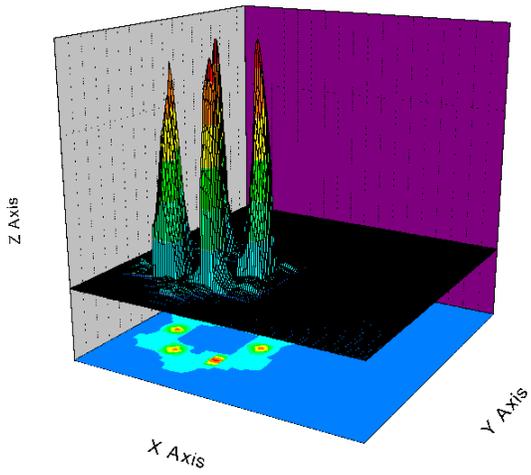

Figure 3(c)

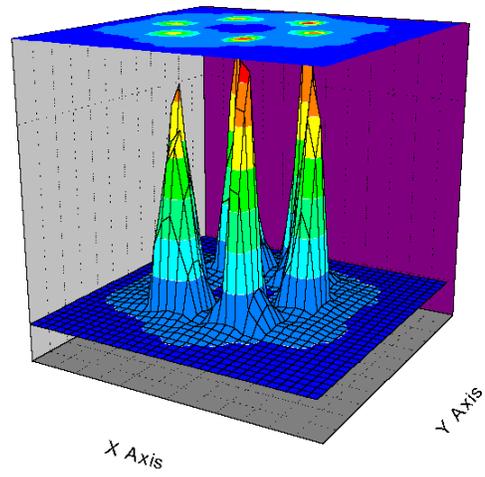

Figure 3(d)

The forces on the nuclei of the ring with LA and pseudoatoms were compared to those of ring in bulk (the benchmark). Figure 4 shows the magnitude of the forces on one of the Si nuclei of the ring resulting from the LA or pseudoatom termination schemes. Since the ring in the rod is in equilibrium, there should not be any forces as is evident from the graph. It is seen that neither the LA nor the pseudo-atom method alone is sufficient to represent the bulk, and additional contributions from the rest of the rod are necessary. These are the longer range Coulomb interactions for which the detailed quantum mechanics is not expected to be important.

Table 1: Comparison of charge densities with various termination schemes

| Distance from plane of ring (in A$^{o}$) | Density in ring in bulk | Normalized difference | |
| --- | --- | --- | --- |
| | | Ring with Link-atoms | Ring with Pseudoatoms |
| Z=0.2 | 942 | 0.05 | 0.02 |
| Z=0.4 | 507 | 0.12 | 0.05 |
| Z=0.6 | 246 | 0.32 | 0.19 |
| Z=0.8 | 171 | 0.42 | 0.42 |

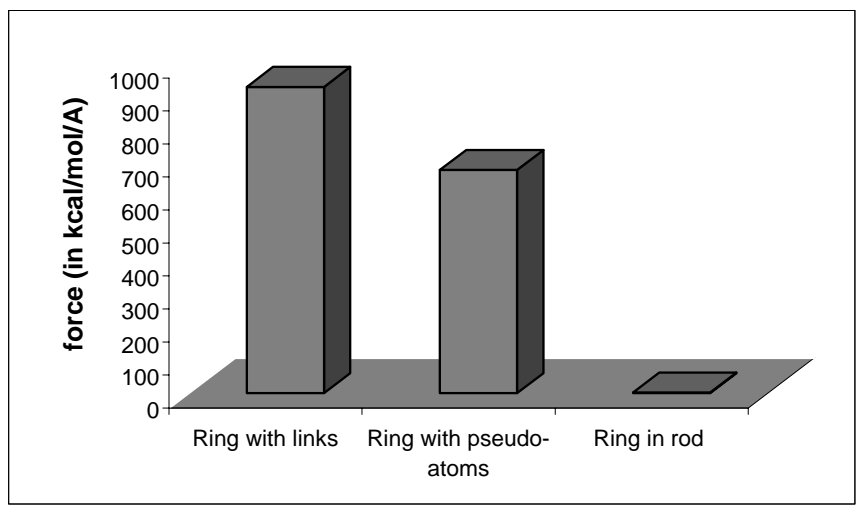

Figure 4: Comparison of forces on Si nuclei with different termination schemes

To improve our embedding procedure, we describe the remainder of the environment as two dipoles for the top and bottom portions of B (Figure 5). The values of the dipole have been calculated using the TH-NDDO charge density for these two portions of the rod. These domains are taken to be charge neutral, but are polarized by the presence of the QM domain A. The validity of the approximation of the rod with dipoles has been checked by finding the difference between the force on Si nuclei due to all the charges and that due to the dipole and its found to be very small of (about 6.25kcal/mol/Å). It is found that even with dipoles placed externally above and below the ring, the force on Si nuclei is 364.7 kcal/mol/Å. Hence, it is clear that using pseudoatoms or external dipoles alone will not lead to a satisfactory embedding.

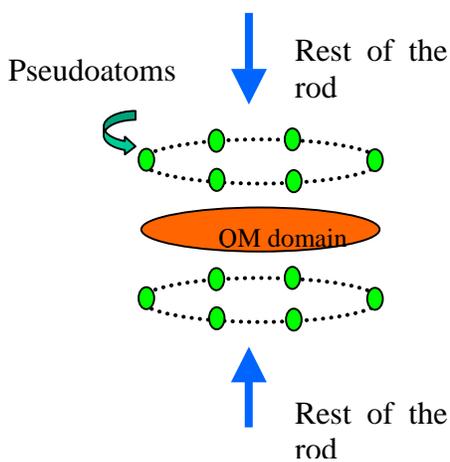

Figure 5: Approximation of the top and bottom portions of the rod by two dipoles

The problem of large forces was solved by including the two dipoles in TH-NDDO calculation to incorporate the effects of polarization of the ring due to the dipoles. Figure 6(a) and 6(b) show the significant improvement in the values forces and charge densities with the incorporation of dipoles respectively. From Figure 6(b), it is seen that the normalized difference of charge density is reduced by an order of magnitude.

Figure 6(a)                                   Figure 6(b)

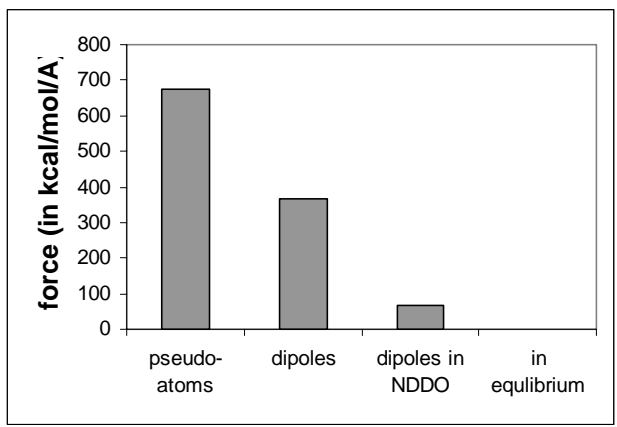 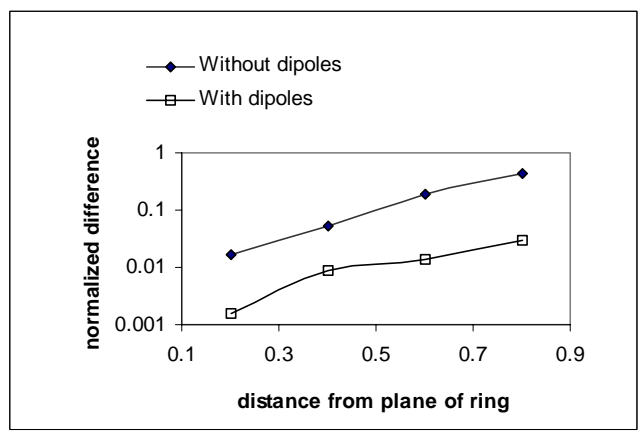

This method was found to be applicable to strained situations and other systems as well. (Table 2 shows all the cases studied.)

Table 2: Comparison of actual forces and charge densities with those obtained from our method, in various cases

| Various Other Cases studied | Comparison of forces (in kcal/mol/A) | | Fractional normalized charge density difference with respect to bulk |
|---|---|---|---|
| | Actual (in presence of bulk) | With pseudoatoms + dipoles | |
| **I. For the rod with 6 rings** | | | |
| a) 5% radially expanded central ring | 214 | 223 | 0.005 |
| b) A distorted central ring | 418 | 437 | 0.002 |
| c) Uniaxially strained rod | 0.5 | 29 | .007 |
| **II. For a longer rod with 10 rings** | | | |
| a) In Equilibrium | 39 | 32 | 0.001 |
| b) 5% radially expanded central ring | 256 | 241 | 0.007 |
| c) A distorted ring | 571 | 639 | 0.005 |
| **III. 3-membered ring** | 17 | 52 | 0.008 |

* in each of above the cases it has been verified that pseudo-atom method is a better termination scheme

As a further test, we have used this method to partition a highly strained small molecule, which is a 3-membered silica ring[21] (Figure 7). These types of ringed structures are found abundantly on surface of amorphous silica. Again, the results are quite satisfactory (see the last row of Table 2).

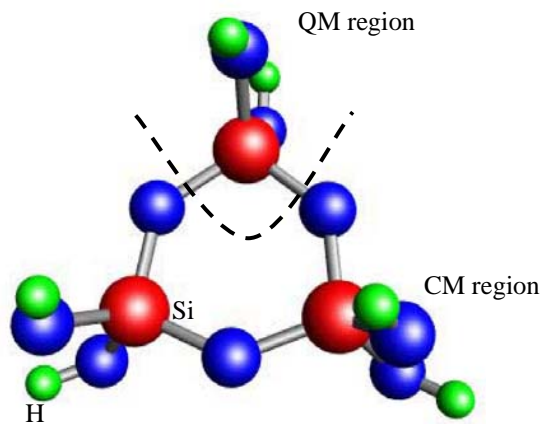

**Figure7 Partitioning of 3-membered silica ring**

**Conclusion and Discussions**

It is seen that the pseudoatoms together with the dipoles in the TH-NDDO can effectively represent the CM region. The generality of the proposed scheme extends to systems such as strained rods and 3-membered cyclic silica ring. Thus, we have succeeded in developing a tool for simulating the exchange interactions and simplifying the environment of the QM region. Furthermore the pseudoatom, in the context of TH, is a better representation of the exchange interactions compared to the LA method.

So far we have calculated the dipole moments from TH-NDDO. This has been possible since the systems under study are small enough to allow full quantum calculations.

For very large systems, like amorphous silica surface, we hope to find a dipole moment associated with Si-O bond from the known dipole moments in these calculations, which then can be included easily in the quantum calculations.

This method provides an example for multi-scale modeling that is internally self-consistent and based at all scales in the fundamental underlying quantum mechanics. Although the study has been limited to Si based materials, it should provide the guidance for extension to other materials as well.

**Acknowledgements**

This work was supported partly by Department of Energy under Grant No. DE-FG03-98DP00218 and by NSF-DMR-0325553. The authors thank Dr. Samuel B. Trickey, Dr. Hai-Ping Cheng for their helpful discussions.

**References**

[1] Field, M.J.; Bash, P. A.; Karplus, M. *J. Comp. Chem.* **1990**, 11, 700

[2] *Combined Quantum Mechanical and Molecular Mechanics Methods*, ACS Symposium Series 712, edited by J. Gao and M. A. Thompson ∼American Chemical Society, Washington, DC, **1998**

[3] Cui, Q.; Guo, H.; Karplus, M. *J. Chem. Phys.* **2002**, 117, 5517.

[4] Bakowies, D.; Thiel, W. *J. Phys. Chem.* **1996**, 100, 10580.

[5] Zhang, Y.; Lee, T. S.; Yang, W. *J. Chem. Phys.* **1999**, 110, 46.

[6] Gao, J.; Xia, X. *Science* **1992**, 258, 631.

[7] Loferer, M. J.; Loeffler, H. H.; Liedl, K. R. *J. Comp. Chem.* **2003**, 24, 1240.

[8] Pu, J.; Gao, J.; Truhlar, D. G. *J. Phys. Chem. A* **2004**, 108, 632.

[9] Kongstead, J.; Osted, A.; Mikkelsen, K. V. *J. Phys. Chem. A* **2003**, 107, 2578.

[10] Warshel, A.; Levitt, M. *J. Mol. Biol.* **1976**, 103, 227.

[11] Singh, U. C.; Kollman, P. A. *J. Comp. Chem.* **1986**, 7, 718.

[12] Assfeld, X.; Rivail, J. L. *Chem. Phys. Lett.* **1996**, 263, 100.

[13] Reuter, N.; Dejaegere, A.; Maigret, B.; Karplus, M. *J. Phys. Chem. A* **2000**, 104, 1720.

[14] Gao, J.; Amara, P.; Alhambra, C.; Field, M. J. *J. Phys. Chem. A* **1998**, 102, 4714.

[15] Maseras, F.; Morokuma, K. *J Comp. Chem.* **1995**, 16, 1170; Dapprish, S.; Kormaromi, I.; Byun, K. S.; Morokuma, K.; Frish, M. J. *J. Mol. Struct: THEOCHEM* 1999, 461-462, 1.

[16] Poteau, R.; Ortega, I.; Alary, F.; Solis, A. R.; Barthelat, J-C.; Daudey, J-P. *J. Phys. Chem.* **2001**, 105, 198.

[17] Taylor, C. E.; Cory, M.; Bartlett, R. J. *Comp. Mat, Sci.* **2003**, 27, 204.

[18] Zhu, T.; Li, J.; Yip, S.; Bartlett, R. J.; Trickey, S. B.; Leeuw, N. *Proceedings Yangtzee Conference* **2002**.

[19] Mallik, A.; Runge, K.; Dufty, J. W. in preparation.

[20] Hsiao, Y-W; Runge, K.; Cory, M. G.; Bartlett, R. J. *J. Phys. Chem. A* **2001**, 105, 704.

[21] Du, M. H.; Cheng, H. P. *J. Chem. Phys.* **2003**, 119, 6418.